\documentclass[fleqn,usenatbib,useAMS]{mnras}
\usepackage{newtxtext,newtxmath}
\usepackage[T1]{fontenc}
\usepackage{amsmath}
\DeclareRobustCommand{\VAN}[3]{#2}
\let\VANthebibliography\thebibliography
\def\thebibliography{\DeclareRobustCommand{\VAN}[3]{##3}\VANthebibliography}

\usepackage[dvipdfmx]{graphicx}	% Including figure files
\usepackage{amsmath}	% Advanced maths commands
\usepackage{color}

\title[Te line in kilonova]{Tellurium emission line in kilonova AT 2017gfo}

\author[K. Hotokezaka et al.]{
	Kenta Hotokezaka,$^{1}$\thanks{E-mail: kentah@g.ecc.u-tokyo.ac.jp}
	Masaomi Tanaka,$^{2}$ Daiji Kato,$^{3,4}$ 
	Gediminas Gaigalas$^{5}$
	\\
	$^1$Research Center for the Early Universe, Graduate School of Science, University of Tokyo, Bunkyo, Tokyo 113-0033, Japan\\
	$^2$Astronomical Institute, Tohoku University, Aoba, Sendai 980-8578, Japan\\	
	$^3$National Institute for Fusion Science, 322-6 Oroshi-cho, Toki 509-5292, Japan\\
	$^4$Department of Advanced Energy Engineering Science, Kyushu University, Kasuga, Fukuoka 816-8580, Japan\\
	$^5$Institute of Theoretical Physics and Astronomy, Vilnius University, Saul\.{e}tekio Ave. 3, Vilnius, Lithuania}

\date{Accepted XXX. Received YYY; in original form ZZZ}

\pubyear{2022}

\begin{document}
\maketitle

		\begin{abstract}
        The late-time spectra of the kilonova AT 2017gfo associated with GW170817  exhibit a strong emission line feature at $2.1\,{\rm \mu m}$.
        The  line  structure develops with time and there is no apparent blue-shifted absorption feature in the spectra, suggesting that this emission line feature is produced by electron collision excitation. We attribute 
        the emission line to  a fine structure line of Tellurium (Te) III, which is one of the most abundant elements in the second r-process peak.  
        By using a  synthetic spectral modeling including fine structure emission lines with the solar r-process abundance pattern beyond the first r-process peak, i.e., atomic mass numbers $A\gtrsim 88$, we demonstrate that [Te III] $2.10\,\rm \mu m$ is indeed expected to be  the strongest emission line in the near infrared region. We estimate that the required mass of Te III is  $\sim  10^{-3}M_{\odot}$, corresponding to
        the merger ejecta of $0.05M_{\odot}$, which is in agreement  with the mass estimated from the kilonova light curve.
		\end{abstract}

\section{Introduction}
The origin of r-process elements is a long-standing problem in astrophysics \citep{Burbidge1957RvMPB,Cameron1957PASP}.
Neutron star mergers have been considered as promising sites of r-process nucleosynthesis \citep{Lattimer1974ApJ}.
 A neutron star merger, GW170817, was accompanied by an uv-optical-infrared counterpart, a kilonova (or macronova) AT 2017gfo, which provides strong evidence that r-process nucleosynthesis occurs in neutron star merger ejecta \citep[see, e.g., ][for reviews]{Metzger2017LRR,Nakar2020PhR,Margutti2021ARA&A}. 
 %The bolometric light curve indicates that a copious amount of r-process elements is produced in this event.   %\citep{Andreoni2017PASA,Arcavi2017,Coulter2017,Cowperthwaite2017,Drout2017,Evans2017,Kasliwal2017,Pian2017,smartt2017Natur,Tanvir2017,Utsumi2017PASJ}. 

A series of spectral data of the kilonova AT 2017gfo was obtained in the the optical and near infrared bands from 0.5 to 10 days after the merger \citep{Andreoni2017PASA,Chornock2017,Kasliwal2017,Pian2017,Smartt2017Natur,Tanvir2017,troja2017Natur}. 
The kilonova AT 2017gfo is dominated by the photospheric emission at the early times. 
The photospheric emission around a few days after the merger peaks in the near infrared band, indicating  the existence of  lanthanides, which have strong absorption at optical to near infrared wavelengths \citep{barnes2013ApJ,kasen2013ApJ,
tanaka2013,Fontes2020MNRAS,Tanaka2020MNRAS,Kawaguchi2018ApJ,Barnes2021ApJ}. 
The early spectra also exhibit several absorption structures including: (i) the $0.8\,{\rm \mu m}$ feature attributed to Sr II or He I \citep{Watson2019Natur,Domoto2021ApJ,Gillanders2022MNRAS,Perego2022ApJ,tarumi2023}  and (ii) the $1.3\,{\rm \mu m}$ and $1.5\,{\rm \mu m}$ features attributed to La III and Ce III, respectively \citep{Domoto2022}. In addition to the elemental identification, \cite{Sneppen2023} demonstrated that the spectra in the photospheric phase are useful to study the geometry of the outer part of the kilonova ejecta, $\gtrsim 0.2c$.

After the photospheric phase, kilonovae enter the nebula phase, where the ejecta is heated by charged decay products of the radioactivity of r-process nuclei and the heat is radiated through atomic emission lines.
Examining kilonova nebular spectra  provides opportunities to identify atomic species synthesized in the merger ejecta that may not appear as absorption lines during the photospheric phase. For instance, \cite{Hotokezaka2022MNRAS} interpreted the  detection of {\it Spitzer} \citep{Villar2018ApJ,Kasliwal2022MNRAS} at $4.5\,{\rm \mu m}$ at $43$ and $74$ days after the merger as  emission lines of selenium (Se) or tungsten (W).
In the early nebular phase, $\sim 10\,{\rm days}$, the infrared  emission is of particular interest  because the absorption opacity due to atomic transitions is lower compared to the optical region \citep[e.g.,][]{Tanaka2020MNRAS}, and thus, the emission lines are expected to appear as early as $\lesssim 10$ days. Most of  infrared emission lines are expected to arise from fine-structure transitions in the ground terms of heavy elements, for which the line wavelengths and transition rates can be obtained with reasonably high accuracy from the experimentally calibrated atomic energy levels. Furthermore, such emission lines can be used to estimate the mass distribution of the emitting ions from the emission line spectra. In fact, the mass distributions of ions in type Ia supernova ejecta have been derived from the infrared nebular spectra \citep[e.g.,][]{Kwok2023ApJ,DerKacy2023ApJ} 
  
%These facts already have been demonstrated in SN 1987A as well as SNe Ia, e.g., 2005dy and 2014J.

In section \ref{sec:1}, we study an emission line feature at $2.1\,{\rm \mu m}$ in the kilonova AT 2017gfo spectra from 7.5 to 10.5 days. We attribute this line  to a fine-structure line of doubly ionized Tellurium (Te III, atomic number 52).   The Te III mass that is required to explain the observed data is estimated as $\sim 10^{-3}M_{\odot}$. With a synthetic spectral modeling with the solar r-process abundance pattern, we show that [Te III] $2.10\,{\rm \mu m}$ is the strongest fine structure emission line in the near infrared region. In section \ref{sec:2}, we conclude the results and discuss the uncertainties and implications.

\section{Te III line in kilonova} \label{sec:1}
The emission lines produced through radiative de-excitation of atoms emerge from the optically thin region of the ejecta.
The optical depth of the kilonova ejecta with an expansion velocity of $v_{\rm ej}$ and a mass of $M_{\rm ej}$ is 
\begin{align}
    \tau & \approx \frac{\kappa M_{\rm ej}}{4\pi (v_{\rm ej}t)^2},\\
    & \approx 1\left(\frac{\kappa}{1\,{\rm cm^2g^{-1}}}\right)
    \left(\frac{M_{\rm ej}}{0.05M_{\odot}}\right)
    \left(\frac{v_{\rm ej}}{0.1c}\right)^{-2}
    \left(\frac{t}{10\,{\rm day}}\right)^{-2},
\end{align}
where $\kappa$ is the opacity and $t$ is the time since merger.
The opacity is dominated by bound-bound transitions of heavy elements and depends on the composition and wavelengths. 
\cite{Tanaka2020MNRAS} show that the expansion opacity decreases with  wavelength, e.g., $\sim 10$ -- $100\,{\rm cm^2g^{-1}}$ around $0.5\,{\rm \mu m}$ and $\lesssim 1\,{\rm cm^2g^{-1}}$ around $2\,{\rm \mu m}$.
Therefore, infrared emission lines are expected to emerge at the earlier time than optical lines. With the ejecta parameters of AT 2017gfo, we expect emission lines to dominate over the photospheric emission as early as $\sim 10\,{\rm days}$ around $2\,{\rm \mu m}$. 

\begin{figure}
\begin{center}
\includegraphics[width=85mm]{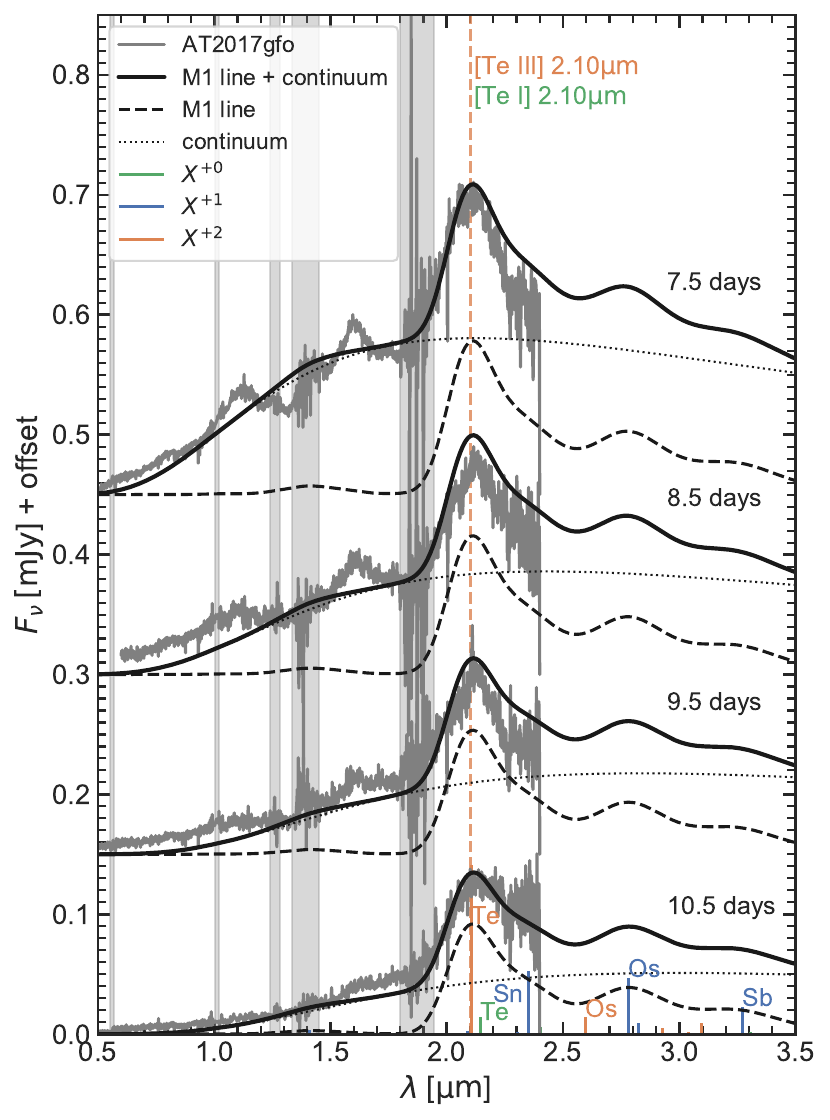}
\caption{Spectral series of the kilonova AT 2017gfo 7.5--10.5 days after the merger. The observed data  were taken by  X-shooter on VLT \citep{Pian2017}.
The synthetic spectra are composed of fine structure emission lines (dashed curve) and 
a continuum (dotted curve), where the continuum emission is approximate by a blackbody with 
temperatures $T_{\rm BB}=2400,\,2100,\,1800,\,1700\,{\rm K}$ at 7.5, 8.5, 9.5 and 10.5 days, respectively.
%$T_{\rm BB}=2400,\,2100,\,1800,\,1700\,{\rm K}$ for Hullac
The electron temperature is fixed to be $2000\,{\rm K}$.
 The ejecta model assumes $M_{\rm ej}=0.05M_{\odot}$,  $v_{\rm exp}=0.07c$,  and $n_e=10^7\,{\rm cm^{-3}}(t/9.5{\rm d})^{-3}$.  We use ionization fractions of $(Y^{+0},\,Y^{+1},\,Y^{+2},\,Y^{\geq+3})=(0.25\,\,0.4,\,0.25,\,0.1$) for all the atomic species for simplicity. The composition is assumed to be the solar r-process abundance pattern with $A\geq 88$ (figure \ref{fig:abundance}).
 %$(Y^{+0},\,Y^{+1},\,Y^{+2},\,Y^{\geq+3})=(0.2.\,\,0.4,\,0.35,\,0.05$) $n_e=2\cdot 10^7\,{\rm cm^{-3}}(t/9.5{\rm d})^{-3}$ for Hullac
 The shape of each emission line is assumed to be a Gaussian profile with a broadening parameter of $0.07c$. 
 The distance to the source is set to $=40\,{\rm Mpc}$. The wavelength of [Te I] $2.10\,{\rm \mu m}$ and [Te III] $2.10\,{\rm \mu m}$ is shown as a vertical dashed line.
 Also shown as vertical lines are possibly strong emission lines at 10.5 day. The gray shaded vertical regions depict the wavelength ranges between the atmospheric windows. The wavelength of [Te I] $2.10\,{\rm \mu m}$ is shown with an offset of $+0.04\,{\rm \mu m}$. 
 %The luminosity in the emission lines is $\sim 1.5\cdot 10^{40}\,{\rm erg/s}$, which accounts for about a half of the total luminosity.   The spectral peak at $2.1\,{\rm \mu m}$ coincides with the wavelength of the fine structure line of [Te III]\,$2.10\,{\rm \mu m}$.  $L_{BB}=5.7,3.3,2.2,1.6\cdot10^{40}$, $L_{M1}=1.2,1.1,0.9,0.9\cdot10^{40}$.
 }
\label{fig:kilonova}
\end{center}
\end{figure}

Figure \ref{fig:kilonova} shows the  spectral series of the kilonova AT 2017gfo from  $7.5$ to $10.5$ days after  the merger taken by X-shooter on the Very Large Telescope \citep{Pian2017}. The observed spectra are composed of several line features  and a continuum component extending from the optical to near infrared bands. We model the underlying continuum spectrum by blackbody radiation, where the photospheric velocity and temperature for $7.5$--$10.5$ days are $0.06$ -- $0.08c$ and $1700$ -- $2400\,{\rm K}$, respectively. 
The observed spectra clearly show
an emission line   at $2.1\,{\rm \mu m}$ (see \citealt{Gillanders2023} for a detailed analysis). The expansion velocity of the line emitting region  is $\sim 0.07c$ derived from Doppler broadening of the line, which is consistent with the picture where the emission line is produced outside the photosphere.
The line flux remains roughly constant with time while the continuum flux declines, and thus,  the line-to-continuum ratio increases from $\sim 1$ at 7.5 days to $\sim 1.5$ at 10.5 days. This development of the emission line without a blue-shifted absorption feature indicates that the emission at $2.1\,{\rm \mu m}$ is a forbidden line driven by electron collision rather than  an emission line associated with an absorption line, e.g., a P-Cygni line or a fluorescence line. %If the ejecta is optically thin to a forbidden line, the flux in the  line is expected to remain roughly constant with time when the  critical density of the forbidden line is lower or comparable to the mean electron density of the ejecta. 
The wavelength of the peak of the emission line feature indeed coincides with a fine structure line, [Te III] $2.10\,{\rm \mu m}$, arising from the  transition between the ground level $^3{\rm P}_0$ and the first excited level $^3{\rm P}_1$. 
It is worth noting that
 [Te III] $2.10\,{\rm \mu m}$ has  been detected in planetary nebulae \citep{Madonna2018ApJ}. Note that the transition between the ground level $^3{\rm P}_2$ and the second excited level $^3{\rm P}_1$ of Te I also produces an emission line at 2.1 ${\rm \mu m}$. {As discussed later, the contribution of Te I line is weaker than Te III line.}

\begin{figure}
\begin{center}
\includegraphics[width=85mm]{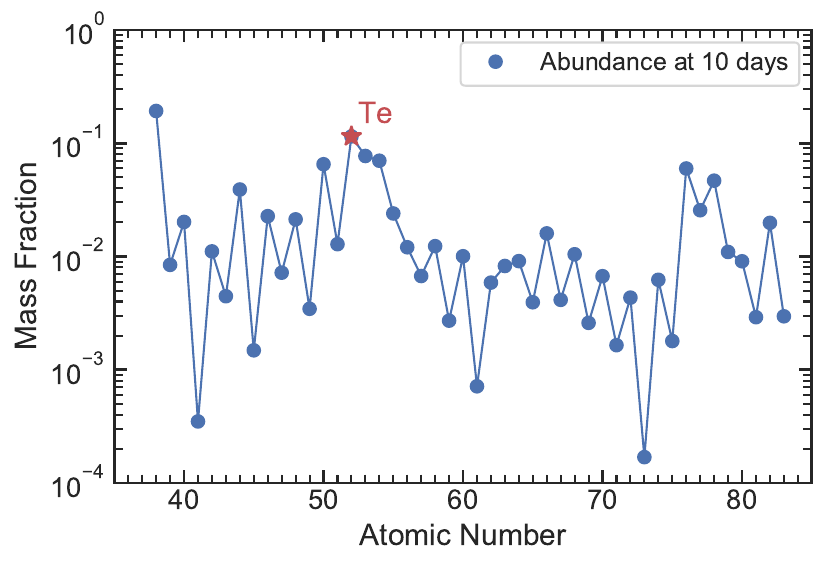}
 \caption{Elemental mass fraction used in the synthetic modeling. The abundance pattern at 10 days is determined such that the final abundance pattern at $\sim 5\,{\rm Gyr}$ matches the solar r-process residual for $88\leq A \leq 209$. Tellurium is the second most abundant element in this model.
 }
\label{fig:abundance}
\end{center}
\end{figure}

It may not be surprising that Te III produces the strongest emission lines  because Te  is among  the most abundant elements in the second r-process peak.
Figure \ref{fig:abundance} shows the mass fraction of each atom at 10 days after the merger. Here we assume that the final abundance pattern matches the solar r-process residual with atomic numbers $A\geq 88$ \citep{Hotokezaka2020ApJ}, i.e., the elements beyond the first r-process peak. With this assumption, the most abundant element is Sr and the second most is Te at 10 days. 
 Note also that [Te III] $2.10\,{\rm \mu m}$ is particularly expected to be strong as long as Te III is abundant outside the photosphere because this line is produced by radiative decay of the first fine structure transition level, which is easily excited by electron collision. For the iron peak elements, [Co III] $11.89\,{\rm \mu m}$ and [Co II] $10.52\,{\rm \mu m}$ represent lines of the same nature. Indeed, these are among the most prominent mid-IR lines observed in SNe Ia and SN 1987A, respectively \citep{Kwok2023ApJ,Wooden1993}.

Let us first give an estimate of the amount of Te III from the observed line flux assuming that the  observed line flux is predominantly produced by Te III and  the ejecta is optically thin to the  [Te III] $2.10\,{\rm \mu m}$ line. The total line luminosity is given by
\begin{align}
    L \sim h\nu_{10} A_{10} f_1 N({\rm Te\,III}), \label{eq:L}
\end{align}
where $h\nu_{10}$, $A_{10}\approx 2\,{\rm s^{-1}}$, and $f_1$ are the excitation energy, the radiative decay rate, and the fraction of Te III ions in the  $^3{\rm P}_1$ level,  respectively, and $N({\rm Te\,III})$ is the total number of Te III ions in the ejecta \citep[see equations (1) and (2) in][ for the formula of M1 transition probabilities]{Hotokezaka2022MNRAS}. The observed flux at $2.1\,{\rm \mu m}$  after subtracting the underlying continuum is $\sim 0.1\,{\rm mJy}$, corresponding to the observed line luminosity of $L_{\rm obs,line}\sim 2\cdot 10^{39}\,{\rm erg\,s^{-1}}$ with $D=40\,{\rm Mpc}$. Equation (\ref{eq:L}) leads to a total mass of Te III:
\begin{align}
    M({\rm Te~III})\sim  10^{-3}M_{\odot} \left(\frac{f_1}{0.1}\right)^{-1}\left(\frac{L_{\rm obs,line}}{2\cdot 10^{39}~{\rm erg~s^{-1}}}\right).
\end{align}
Note that the electron density at $t\sim 10.5$ day may be comparable to
 the critical density of Te III $^3{\rm P}_1$ \citep{Madonna2018ApJ}, and therefore, the level fraction $f_1$ is comparable to or slightly less than that expected from the thermal distribution, i.e., $f_{1}\approx 0.1$ {in thermal equilibrium at} $T_e=2000\,{\rm K}$.
Given the total ejecta mass of $\sim 0.05M_{\odot}$, we estimate that the mass fraction of Te is greater than a few per cent.% if Te III dominates the ionization state \citep[see, e.g., ][for study of ionization in kilonova nebulae]{Hotokezaka2021MNRAS,Pognan2022MNRAS}.

We now turn to the comparison of the observed spectra with a synthetic spectral model. The synthetic spectrum is composed of fine structure emission lines and a continuum component, where the continuum emission is  approximated by blackbody radiation. The blackbody temperature and radius at a given epoch are determined such that the synthetic spectrum roughly agrees with the observed one at the near IR region $\lesssim 2\,{\rm \mu m}$. 
The emission line spectrum is computed by the one-zone modeling presented in \cite{Hotokezaka2022MNRAS},
where the energy level populations are solved by balance between collision and radiative decay for a given electron density, temperature, and ionization state. We use the collision strengths of the fine structure transitions of the ground term of Te III derived by \cite{Madonna2018ApJ}. The collision strengths of other elements that are relevant for the nebular modeling at $\lambda \lesssim 3.5\,{\rm \mu m}$
are obtained by using an atomic structure code \texttt{Hullac} (\citealt{Bar-Shalom2001JQSR}, see also \citealt{Hotokezaka2022MNRAS}) and
the M1 line list is constructed by using the NIST database \citep{NIST_ASD} and the LS selection rules with the single-configuration approximation (Hotokezaka et al. in prep). Note that the wavelengths and radiative transition rates of the M1 lines in the list are sufficiently accurate for our purpose. 

In the  modeling, the ejecta composition is assumed to be the solar r-process abundance pattern with $A\geq 88$ (figure \ref{fig:abundance}), which is the same as the second and third peak model used in \cite{Hotokezaka2022MNRAS}. 
The model also assumes the electron temperature, $T_e=2000\,{\rm K}$ and the ionization fractions $(Y^{+0}, Y^{+1}, Y^{+2}, Y^{\geq+3})= (0.25, 0.4, 0.25, 0.1)$\footnote{We neglect the emission lines of ions in $Y^{\geq+3}$.}. {These quantities are assumed to be constant with time for simplicity.}
This choice of the ionization fraction is somewhat motivated by \cite{Hotokezaka2021MNRAS}, where the ionization fractions of Nd atoms in the kilonova nebular phase are studied. With these parameters, the ejecta mass of $0.05M_{\odot}$ and the expansion velocity of $0.07c$, 
{[Te III] $2.10\,{\rm \mu m}$ is the strongest emission line among M1 transitions of all the heavy elements beyond the first r-process peak and}
the synthetic spectra can roughly reproduce the emission line structure around $2.1\,{\rm \mu m}$. However, one should keep in mind that the ionization fraction  varies among different atomic species.
The ejecta mass of $0.05M_{\odot}$  agrees with the ejecta mass estimated from the energy budget of the bolometric light curve \citep{Waxman2019ApJ,Kasen2019ApJ,Hotokezaka2020ApJ}.

If this interpretation is correct, we expect that the $2.1\,{\rm \mu m}$ line remains at the later times while the continuum flux keeps declining.
It is worth noting that
Te III may produce another emission line  at $2.93\,{\rm \mu m}$ arising from the transition between the first and second excited levels ($^3{\rm P}_1$-$^3{\rm P}_2$) at the later times because the electron temperature is expected to gradually increase with time   \citep{Hotokezaka2021MNRAS,Pognan2022MNRAS}. Although this line may be hidden by several other lines of Os II, III, and Pd III,
detecting the two lines of Te III  in future events can provide solid confirmation of the Te III production in mergers.
Furthermore,  the ratio of these line fluxes can be used to diagnose the electron temperature.

\section{Conclusion and discussion}\label{sec:2}
The observed spectra of the 
kilonova AT 2017gfo exhibit a strong emission line at $2.1\,{\rm \mu m}$. The  emission line with the lack of an apparent blue-shifted absorption feature suggests that the emission feature is a forbidden line excited through electron collision. We attribute this line to the fine structure line, [Te III] $2.10\,{\rm \mu m}$, which has also been detected in planetary nebulae \citep{Madonna2018ApJ}. Note that Te is one of the most abundant elements in the second r-process peak. We estimate that the mass of Te III is roughly $ 10^{-3}M_{\odot}$ to account for the observed line flux.

We compare the observed spectra with a synthetic model, where the spectrum is composed of fine structure emission lines and a continuum component approximated by blackbody radiation. The spectrum of fine structure emission lines is computed with the one-zone model presented in \cite{Hotokezaka2022MNRAS}.
{With the solar r-process abundance beyond the first r-process peak, $T_e\sim 2000\,{\rm K}$, and $Y^{+2}\sim 0.3$, we show that [Te III] $2.10\,{\rm \mu m}$ is the strongest emission line among M1 transitions of all the heavy elements around 10 days after merger.}
Our model agrees with the observed spectra for the ejecta of $0.05M_{\odot}$ with the solar r-process abundance pattern with $A\geq 88$, an expansion velocity of $0.07c$, and electron temperature of $\sim 2000\,{\rm K}$, and the ionization fraction of $Y^{+2}\sim 0.3$. Because  blackbody radiation may be a poor approximation to the continuum flux around $2\,{\rm \mu m}$ we should keep in mind that the amount of Te III in our analysis may be affected by the continuum flux model. It is also interesting to note that the same abundance pattern can reproduce  the {\it Spitzer} $4.5\,{\rm \mu m}$ detection at 40 days \citep{Villar2018ApJ,Kasliwal2022MNRAS}, in which the emission is attributed to a fine structure line of W III \citep{Hotokezaka2022MNRAS}. {Note that, if the lighter r-process elements are abundant, they are expected to produce emission lines around $2\,{\rm \mu m}$ such as [Kr III] $2.20\,{\rm \mu m}$ and [Se IV] $2.29\,{\rm \mu m}$. However, the observed spectra peak at $2.1\,{\rm \mu m}$, suggesting that these ions are less abundant compared to Te III in the line emitting region of the ejecta.}

Our model does not include electric dipole (E1) lines, which may produce strong absorption and emission lines. 
Recently, \cite{Gillanders2023} suggested that $2.1\,{\rm \mu m}$ may be composed of two lines and an E1 line of Ce III is the best candidate producing this line feature.  
Although we cannot quantify the contamination of E1 lines to the $2.1\,{\rm \mu m}$ feature, we emphasize that the M1 emission of Te can account the observed line flux with reasonable parameters. 
To verify this hypothesis we need spectral modelings with E1 lines. 
%and kilonova spectra from space, which are not affected by the atmospheric absorption. 
We also note that the M1 lines in our list cannot account for the observed feature at $1.6\,{\rm \mu m}$. The flux in this line declines with time as  the continuum flux declines, indicating that this emission feature may be produced by E1 lines. Interestingly, \cite{Domoto2022} show that Ce III has several strong E1 lines around $1.6\,{\rm \mu m}$.

From the early blue emission in the photospheric phase, 
it is suggested that the emission is dominated by the ejecta 
composed of light r-process elements in order to avoid significant absorption
in the optical band by lanthanides. Furthermore, the analyses of the kilonova spectra in the photospheric phase lead to the similar conclusion.  The absorption feature around $0.8\,{\rm \mu m}$ is likely caused by one of light r-process elements, Sr ($Z=38$), or even  He  \citep{Watson2019Natur,Gillanders2022MNRAS,tarumi2023}. \cite{Domoto2022} propose that La ($Z=57$) and Ce ($Z=58$) produce the absorption lines around $1.2$ and $1.5\,{\rm \mu m}$, respectively. But the abundances of La and Ce inferred from the spectral analysis are lower than the solar r-process residuals by factor of $\sim 10$. These indicate that  the outer part of the ejecta ($v\gtrsim 0.2c$) is predominantly composed  of light r-process elements.
In contrast to the early emission, our analysis implies that heavier elements, i.e., the second r-process peak, are likely more abundant in the slower part of the ejecta.

In order to obtain better constraints on the elemental 
abundances and ejecta parameters, the spectral modelings should be improved by developing non-LTE radiation transfer modelings \citep[e.g.,][]{Pognan2022MNRASopacity} and by improving atomic data such as the radiative transition rates \citep[e.g.,][]{Gaigalas2019ApJS}, collision strengths, and recombination rate coefficients.
For future kilonova events, the spectroscopic observations with the JWST as well as ground-based telescopes will be useful to identify more elements in the nebular spectra with a wider wavelength range.

\section*{Acknowledgments}
We thank Nanae Domoto and Yuta Tarumi for useful discussion. This research was supported by JST FOREST Program (Grant Number JPMJFR212Y, JPMJFR2136), NIFS Collaborative Research Program (NIFS22KIIF005), the JSPS Grant-in-Aid for Scientific Research (19H00694, 20H00158, 21H04997, 20K14513, 20H05639, 22JJ22810). 

\section*{DATA AVAILABILITY}
The data presented this article will be shared on  request to the corresponding author.

\bibliographystyle{mnras}
\bibliography{ref.bib}
\end{document}